# An Eclipse Plugin to Support Code Smells Detection

Tiago Pessoa[1], Fernando Brito e Abreu[2,1],
Miguel Pessoa Monteiro[3,1], Sérgio Bryton[1]

[1] CITI/FCT/UNL, Campus da Caparica, 2829-516 Caparica, Portugal
[2] ISCTE-IUL, Av.ª das Forças Armadas, 1649-026 Lisboa, Portugal
[3] DI/FCT/UNL, Campus da Caparica, 2829-516 Caparica, Portugal
{tap16004@fct.unl.pt, fba@iscte.pt, mmonteiro@di.fct.unl.pt, bryton@di.fct.unl.pt}

**Abstract.** Eradication of code smells is often pointed out as a way to improve readability, extensibility and design in existing software. However, code smell detection in large systems remains time consuming and error-prone, partly due to the inherent subjectivity of the detection processes presently available. In view of mitigating the subjectivity problem, this paper presents a tool that automates a technique for the detection and assessment of code smells in Java source code, developed as an Eclipse plug-in. The technique is based upon a Binary Logistic Regression model and calibrated by expert's knowledge. A short overview of the technique is provided and the tool is described.

## 1 Introduction

As advocated by the agile XP methodology [1], refactoring techniques are sought to reduce costs associated with software life cycle at both the Construction phase [2] and the Production phase [2] by supporting iterative and incremental activities and also by improving software extensibility, understandability and reusability [3]. Taking into account that software maintenance activities are the most costly in the software life cycle [4-6], tangible benefits are expected from regularly performing refactoring. Empirical evidence showing the dire consequences of code infested with smells, seems to concur [7].

Even with an approach based on guidelines offered by Beck [1] and Fowler [3], the need of informed human assistance is still felt, to decide where refactoring is worth applying [8]. It is here that the concept of *code smells* provides a contribution [3]. Nevertheless, we have found, in the context of post-graduate courses, that the manual detection of code smells is an excessively time-consuming activity (therefore costly) and is error-prone, as it depends on the developer's degree of experience and intuition.

Empirical studies on the effectiveness of code smells detection techniques are still scarce, but there is some evidence that their eradication is not being achieved to a satisfactory degree, often because developers are not aware of their presence [9]. This is due to the lack of adequate tool support, which requires sound techniques for code smells diagnosis. The subjective nature of code smells definition hinders that soundness [3, 10].

Currently used code smells detection techniques come in two flavours. The first concerns qualitative detection using (inevitably biased) expert-based heuristics. The latter uses thresholds on software metrics obtained from the source code under analysis and seems more appealing for supporting automation due to its repeatability. However, it has two important preconditions for effective use. First, the same set of metrics cannot be used to detect all smells of a catalog such as the one in [3] since code smells are very distinct in nature. Second, even with a customized set of metrics chosen by an expert for detecting a particular smell, the resulting model must be calibrated, i.e., its internal values must be determined to reduce false positives and false negatives. That entails an empirical validation based on existing classification data. Mantyla el al. [11] confirm the difficulty of assessing code smells by using metric sets and the hard task of defining a detection model.

Our work contributes to the field of code smells detection by providing an automated process, supported by a tool (an *Eclipse* plug-in), capable of code smell assessment in Java source code in an objective and automatic way. In contrast with existing proposals that rely purely on the opinion of a single expert, we propose a statistically based detection algorithm that will go through progressive calibration based upon a developers' community. The detection algorithm, based on *Binary Logistic Regression*, was initially calibrated by using a moderately large set of pre-classified methods (by human experts) and validated for the *Long Method* code smell, as depicted in Bryton et al. [10]. The larger the set, the better will be the detection. Our approach relies on the community of our tool users to perform continuous recalibration of the code smells detection models (one per each smell).

We have developed a prototype version of the *Smellchecker* tool, an Eclipse plugin for detecting code smells in Java code. This prototype allows smell tagging, visualization and detection.

The rest of this paper is structured as follows. Section 2 overviews the Binary Logistic Regression model and discusses how calibration using expert's knowledge leverages it. Section 3 introduces the automated process to code smells detection. Section 4 describes the architecture of the *Smellchecker* Eclipse plug-in in detail and summarizes how it is used. Section 5 depicts threats to validity. Section 6 briefly reviews related work. Finally, section 7 presents some closing remarks and outlines future research directions.

## 2 Binary Logistic Regression

Binary logistic regression (BLR) is used for estimating the probability of occurrence of an event (here, the existence of a code smell) by fitting data to a logistic curve. It is a generalized linear model where the dependent variable has two values (code smell present or absent) and an arbitrary set of numeric explanatory variables can be used (here, a set of code complexity metrics). The following logistic function is used to estimate the percentage of probability of a particular code smell:

$$f(z) = \frac{1}{1+e^{-z}} \quad \wedge \quad z = \beta_0 + \beta_1 \times x_1 + \beta_2 \times x_2 + \ldots + \beta_n \times x_n.$$

Where $z$ is called the logit, $x_k$ are the regressors or explanatory variables (code complexity metrics collected from the source code) and $\beta_k$ are the regression coefficients calculated during the calibration process. The choice of the adequate metrics to select for each code smell estimation model based on BLR can be performed by using the Wald or the Likelihood-Ratio tests.

To perform BLR calibration with a statistical tool such as *SPSS* or *R*, we need a sample with values for all variables (explanatory and outcome). Table 1 presents an extract of such a sample, corresponding to four methods on the *org.apache.commons.cli* package from *Apache Commons CLI 1.2*. The collected metrics are **MLOC** (method lines of code), **NBD** (nested block depth), **VG** (cyclomatic complexity), **PAR** (number of parameters) and **LVAR** (number of local variables). These are the explanatory variables in the BLR model. **Long Method** is the dependent or outcome variable: an expert indication of the presence of the *Long Method* code smell on the particular method.

After calibration and validation of the regression coefficients, the instantiated model is used to predict the possible presence of a particular code smell.

**Table 1.** Sample extract for calibrating a *Long Method* code smell estimation model

| Application | ApacheCommonsCLI1.2 | | | |
|---|---|---|---|---|
| **Package** | org.apache.commons.cli | | | |
| **Class** | GnuParser | Parser | HelpFormatter | PosixParser |
| **Method** | flatten | parse | renderOptions | burstToken |
| **MLOC** | 69 | 67 | 59 | 46 |
| **NBD** | 5 | 5 | 4 | 4 |
| **VG** | 11 | 14 | 10 | 6 |
| **PAR** | 3 | 4 | 5 | 2 |
| **LVAR** | 9 | 12 | 19 | 5 |
| **Long Method** | 0 | 1 | 1 | 0 |

## 3 Automated Code Smells Detection

This section sets the context and presents this work's underlying main theme: reducing the subjectivity in code smells detection by automating its process. Figure 1 outlines the automated process, which comprise the activities described next.

**Code Annotation.** In the first iteration, experts must tag the code sample for the presence of code smells in methods, classes or interfaces, to yield an adequate sample for the initial calibration of the models, prior to making the tool available to "regular" developers. In subsequent iterations those developers will only tag false positives (developer disagrees with a detected smell) and false negatives (developer identifies a non-detected code smell). These cases are expected to decrease over time as more data results in more finely calibrated models that produce more precise results.

**Metrics Calculation.** Automatic process that requires a parser-enabled tool that computes metrics on the target source code (the one that is annotated).

**Models Calibration.** Calibration of the BLR models by calculating and validating the regression coefficients. It is an automatic process performed by a statistical processor. Note that there will be one model for each code smell. Each model may have different explanatory variables (metrics).

**Smells Detection.** Application of the calibrated BLR models to selected source code elements. This estimates the probability of presence of the corresponding code smells in the selected artifacts.

**Smells Visualization.** Identification of the source code artifacts where code smells are estimated to be present. The developer can set the threshold probability (e.g. see only the code smells above 90% probability) independently for each code smell.

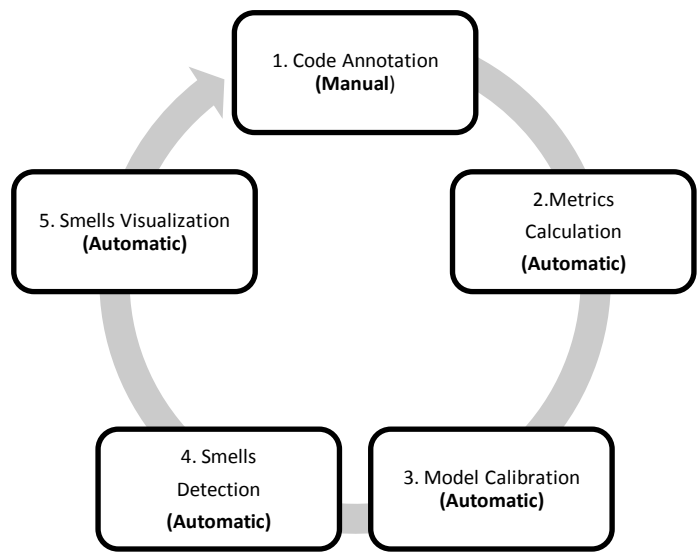

**Fig. 1.** Code smells detection process

Detected code smells will vary depending on the selected probability threshold. Increasing the probability too much will cause more false negatives, while decreasing it in excess will cause more false positives. It will be up to the developer to fine tune the threshold to get the adequate level of advice (let us call it "sensitivity") regarding the presence of code smells. It will also be up to the developer to decide on the adequacy of applying a given refactoring to remove a detected code smell.

The described process can have two different usage patterns. The first concerns a single user with a single machine. The second, a remote usage for more than one user.

**Local Usage.** The process is completely local (Fig. 2). The models calibration process is done locally. The user is responsible for tagging an initial source code base to calibrate the models. Then, the user can apply the models to detect the occurrence of code smells in all code bases of his choice. It is also possible to refine the calibration of the model by providing additional code smell tagging information.

The usefulness of this option is one of practical value: tuning the models, through progressive calibrations, to personal user preferences, thus matching the models to the user notions of where a code smell might be present.

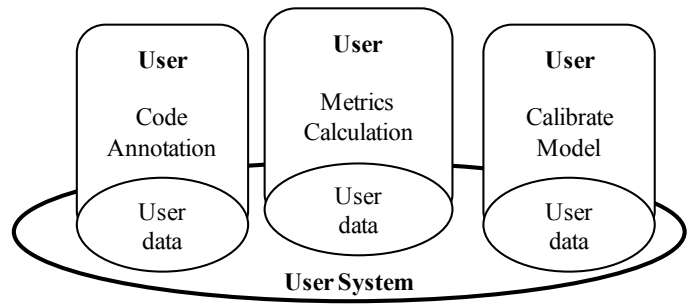

**Fig. 2.** Local usage

**Remote Usage.** The process has a remote central server responsible for storing, on its own data base, the code smells tagging and metrics values provided by several users (Fig. 3). With the calibration and validation of the BLR model being performed on the server, users can remotely query the server for the most recent model parameters. This model is calculated from the aggregated data provided by all users, augmenting the statistical significance of the BLR estimates and thus providing a more accurate detection of the code smells.

One of the simpler updates the user can make is to provide feedback on false positives and false negatives detected, thus contributing for the models' progressive enhancement.

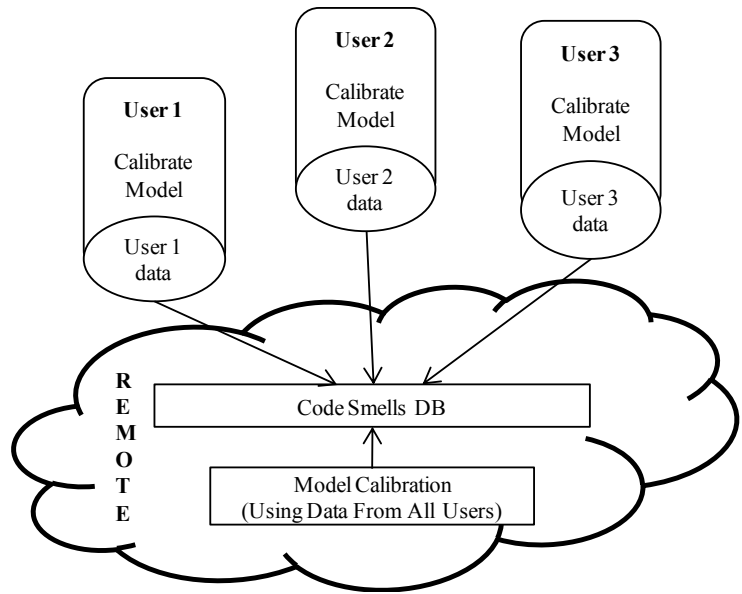

**Fig. 3.** Remote usage

## 4 Smellchecker

We have chosen the *Eclipse* framework as the target platform to support *Smellchecker* development due to its advanced Java support, available refactoring features, along with its plug-in development facility,

Eclipse is a stylish and appropriate choice for our tool deployment since its architecture by components supports integration of virtually any component within its architecture. Yet, despite its advanced support for Java source code refactoring as part of its standard JDT toolkit, code smells detection is by and large completely lacking.

Since Eclipse's architecture by components permits seemingly integration of virtually any component within its scheme, it is a stylish and appropriate choice for our tool deployment.

The *Smellchecker* prototype architecture uses Java 1.6 and Eclipse platform 3.5. It comprises the following components, traceable to the processes described in Fig 1:

**1. Source Code Annotation** – Eclipse's SWT/JFace UI Framework, Eclipse's JDT AST;
**2. Metrics Calculation** – Metrics Eclipse Plugin Version 1.3.8;
**3. BLR Models Calibration** – R Statistical Computing, JRI;
**4. Code Smells Detection** - Eclipse's JDT AST;
**5. Code Smells Visualization** - Eclipse's SWT/JFace UI Framework.

An overview of all the major components comprising *Smellchecker* is provided in Fig. 4.

Common to all processes is the persistence component, represented on the diagram as the Code Smells DB (database) component.

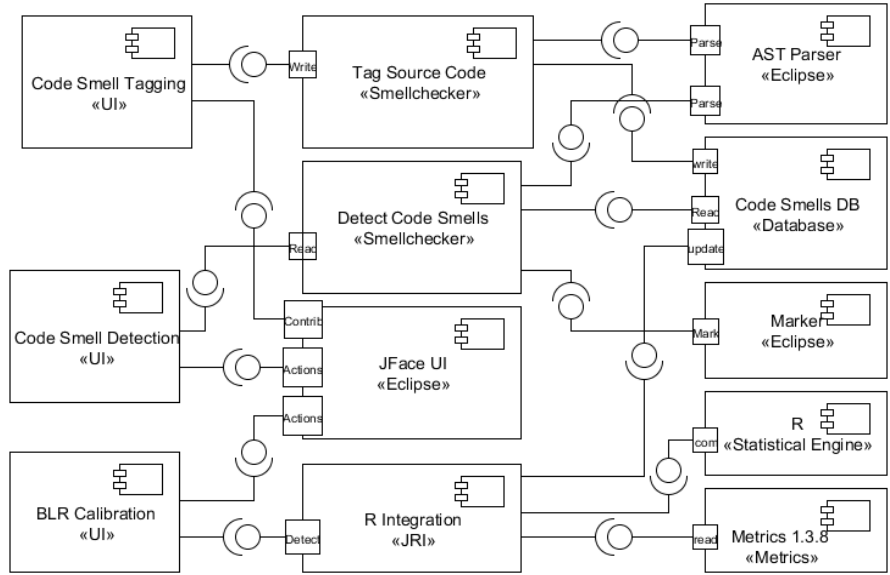

**Fig. 4.** Smellchecker component diagram

**Source Code Annotation.** Eclipse's SWT/JFace UI Framework provides user interface resources that allow code smells tagging assistance.

Users can tag classes, methods and interfaces. This operation may be performed manually, by tagging the code directly with Java annotations with the following syntax:

```
@CodeSmell(type=CodeSmellType.LargeClass,
          description="Too many functionalities")
public class Customer {...}
```

The code smells annotation process is also assisted by the UI. The user can select from a drop menu the corresponding code smell tag for each desired code fragment. Eclipse will present a *Tag Smell* menu in the context of code elements when available. As described by Fowler [3] and Wake [12], classes, interfaces and methods have particular code smells associations. An example of the UI assisted code smell annotation process in represented in Fig 5.

After selection, the annotation will be added, via AST (Java's Abstract Syntax Tree) API, to the source code exactly as the annotation example listed above. A data base entry will be made consisting of the smell indication in conjunction with information of the method, class, interface, application, and package annotated.

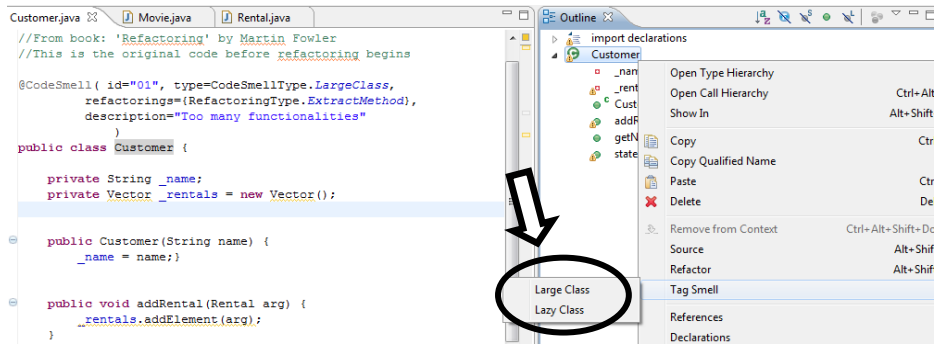

**Fig. 5.** UI assisted Code Smell Annotation

Fig. 6 shows the command and icon, *Refresh Visualization*, for constructing or refreshing source code's smells annotation database. The information will then be displayed in tree form on the workbench view *Smellcheckeer Tagged Code Smells* in conjunction with the metrics calculated for that specific resource. This view is context aware so a class, package, or method must be selected.

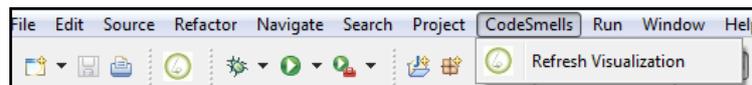

**Fig. 6.** Refresh Visualization command and icon

**Code Metrics Calculation.** Metrics calculation is accomplished by the Eclipse Metrics 1.3.8 plug-in. Metrics supported for classes include *Lines of Code*, *Depth of Inheritance Tree*, *Number of Methods* and *Lack of Cohesion* to name a few. For methods, the tool supports: *Lines of Code*, *Nested Block Depth*, *McCabe Cyclomatic Complexity* and *Number of Parameters*. It also provides an extension point on the basis of which it is possible to extend the plug-in to calculate additional metrics.
The Metrics Calculation plugin must be enabled via the *Smellchecker preference page* menu. Metrics are then calculated during the build cycle and displayed in the *view Smellchecker: Tagged Code Smells* in conjunction with code smells presence indications. Fig. 7 shows an example.

| Metrics | Total |
|---|---|
| Number of Overridden Methods | 0 |
| Number of Attributes | 2 |
| Number of Children | 0 |
| Method Lines of Code | 38 |
| Number of Methods | 4 |
| Depth of Inheritance Tree | 1 |
| Lack of Cohesion of Methods | 0,66667 |
| Number of Static Methods | 0 |
| Specialization Index | 0 |
| Weighted methods per Class | 12 |
| Number of Static Attributes | 0 |

**Fig. 7.** Metrics and Annotated Code Smells view

**Models calibration.** Calibration and validation of the BLR models is performed by the R statistical engine. Interaction between the plug-in and R is made with JRI, a Java/R Interface that allows running an instance of R as a process that responds to command line type commands and outputs back to Java the data resulting from its computations. Actions that must be performed include: normality tests on the metric data (to know which test to apply), correlation coefficients among the variables, collinearity diagnosis tests and goodness-of-fit analysis for the BLR models.

**Smells Detection.** It follows the successful calibration of the BLR model. The selection of the command *Code Smells Detection*, on the toolbar *CodeSmells* menu, parses (for all open projects) all compilation units with the AST API parse command of JDT. For each class or method (depending if the model was calibrated for a class or a method smell), the corresponding metrics is read from the data base and the probability of the presence of the code smell is calculated. If the probability is above the defined threshold, the node of the AST corresponding to the artifact under analysis is marked as a problematic node with JDT' Marker class, which is responsible for marking compile errors or warnings. To distinguish make code smells warnings from compile errors and warnings, a lower value of importance is linked to the node where the code smell was detected.

**Smells Visualization.** Since nodes identified by the model have been marked, they will appear in the error log view and they will have the same properties as compiler errors. So a jump to the smelly section of the code can be performed upon a click.

The previous descriptions apply both to the single-user/single-machine functioning, as well as to the remote usage process. Additional conditioning pertains to the centralized efforts of the server, as well as availability of the service as a web service to accommodate client needs for a simple interface. The data is communicated as an

XML document and the derivation of the model is identical with the difference that in the server case, more data is expected to be available.

## 5 Threats to Validity

Assurance cannot be stated that, for all the 22 Code Smells described by Fowler, the *BLR* model proposed will derive valid assumptions. For example, *Duplicated Code* is an active area of research with sophisticated mechanisms already derived to identify Code Clones [13].

Threats to validity arise from the set of metrics selected, the experts opinion and the sensibility expressed by the BLR model after calibration.

Metrics selected for the *Smellchecker* tool derive from their availability in a prior existing Eclipse Plugin and represent a small subset of metrics presented in literature [14]. Metrics in *Smellchecker* concern classic complexity measures that are not suited to detect all code smells. Therefore the need exist to extend their boundary.

Further studies must be conducted to know how sensitive the BLR model is to bad data given by experts and also to analyze over fitting issues and scalability of the model.

## 6 Related Work

A few open-source tools exist for detecting code smells in Java code. Most of them use static analysis, that is, they do not require executing the program, such as the one presented in this paper.

**PMD (**http://pmd.sourceforge.net/**).** This widely used tool uses static analysis techniques to scan Java source code and look for potential problems like *possible bugs* (empty try/catch/finally/switch statements), *dead code* (unused local variables, parameters and private methods), *suboptimal code* (wasteful String/StringBuffer usage), *overcomplicated expressions* (unnecessary *if* statements, *for* loops that could be *while* loops) and *duplicate code* (copied/pasted code means copied/pasted bugs). PMD is integrated with *JDeveloper, Eclipse, JEdit, JBuilder, BlueJ, CodeGuide, NetBeans, IntelliJ IDEA, TextPad, Maven, Ant, Gel, JCreator,* and *Emacs*.

**FindBugs (**http://findbugs.sourceforge.net**).** This tool is also widely used and integrated with *Eclipse*, using the static analysis capabilities of Apache's Byte Code Engineering Library (BCEL) to inspect Java bytecode for occurrences of bug patterns. The latter are code idioms that are often errors. Bug patterns arise for a variety of reasons such as: difficult language features, misunderstood API methods, misunderstood invariants when code is modified during maintenance, or garden variety mistakes as typos or use of the wrong boolean operator. Their authors report that its analysis is sometimes imprecise since many false positives (up to 50% of identified bugs) can be risen.

**SISSy (http://sissy.fzi.de).** According to its authors, the Structural Investigation of Software Systems tool can detect some well-known code smells and the violation of over 50 typical OO design principles, heuristics and patterns, such as bottleneck classes, god classes, data classes or cyclical dependencies between classes or packages. SISSy can analyze systems written in Java, C++ or Delphi but, as far as we could ascertain, is not integrated with any IDE.

**Smelly (http://smelly.sourceforge.net).** Is an Eclipse plug-in that, according to its authors, is able to detect the following code smells in Java code: *Data Class, God Class, God Method, High Comment Density,* **Long Parameter List** and *Switch*. Only the one in bold matches the original name in the original code smells catalog [3].

**Code Bad Smell Detector (http://cbsdetector.sourceforge.net/).** This tool claims to detect five of Fowler et al. [3] code smells: *Data Clumps, Switch Statements, Speculative Generality, Message Chains,* and *Middle Man*, from Java source code. It has no recent downloads and appears to be associated with an ongoing PhD work. It is also not integrated with an IDE.

## 7 Conclusion & Future Work

**Closing Remarks.** The idea of automating code smells detection by using metrics and tools is not new. However, the detection technique used in the *Smellchecker* tool is in contrast with all other known proposals due to the usage of a dynamic statistical process that relies on expert's knowledge that can be applied, theoretically[1], to any smell. A distinctive characteristic of our approach is that the quality of the detection process increases as time goes on, due to progressive model calibrations supported by accumulated data.

**Future Work.** More in-depth studies are required for validating the process for different code smells, different calibration data, and better assurance of the *BLR* model coefficients.

According to [3], a given code smell can be mitigated/removed by applying one out of a set of refactoring transformations. Since *Eclipse* supports several of those transformations, we envision that upon code smells identification, adequate refactorings could be suggested to remove the smell. In future, we will look at ways of computing the expected quality improvements attained by applying each of the refactoring alternatives. Hopefully, that will allow us to provide some advice for the developer.

---

[1] Empirical studies are required to validate this assumption.

## Acknowledgements

The work presented herein was partly supported by the VALSE project of the CITI research center within the Department of Informatics at FCT/UNL in Portugal.